\documentclass[%
 reprint,
superscriptaddress,
amsmath,amssymb,
aps,
floatfix,
]{revtex4-2}

\usepackage{graphicx}
\usepackage{dcolumn}
\usepackage{bm}
\usepackage{braket}%
\usepackage{here}
\usepackage[colorlinks=true, allcolors=blue]{hyperref}
\usepackage{color}
\usepackage{textcomp}

\begin{document}

\preprint{APS/123-QED}

\title{Difference in peripherality of the inclusive $(p,p'x)$ and $(d,d'x)$ reactions and its implications for phenomenological reaction model}

\author{Hibiki Nakada}  %
\email[Email address: ]{nakada27@rcnp.osaka-u.ac.jp}
\affiliation{Research Center for Nuclear Physics, Osaka University, Ibaraki, Osaka 567-0047, Japan}
\author{Shinsuke Nakayama}  %
\affiliation{Nuclear Date Center, Japan Atomic Energy Agency, Tokai, Ibaraki 319-1195, Japan}

\author{Kazuki Yoshida}  %
\affiliation{Advanced Science Research Center, Japan Atomic Energy Agency, Tokai, Ibaraki 319-1195, Japan}

\author{Yukinobu Watanabe}  %
\affiliation{Department of Advanced Energy Science and Engineering, Kyushu University, Fukuoka 816-8580, Japan}
\author{Kazuyuki Ogata}  %
\affiliation{Department of Physics, Kyushu University, Fukuoka 819-0395, Japan}
\affiliation{Research Center for Nuclear Physics, Osaka University, Ibaraki, Osaka 567-0047, Japan}

             

\begin{abstract}
\noindent
\textbf{Background}:
Previous studies have revealed the importance of introducing surface correction into a phenomenological model for inclusive $(n,n'x)$ and $(p,p'x)$ reactions.
These findings have contributed significantly to the improvement of nuclear data evaluation.
However, the necessity for the surface correction in an inclusive $(d,d'x)$ reaction has hardly been investigated.

\noindent
\textbf{Purpose}:
The purpose of this study is to investigate the difference in the peripherality of the $(p,p'x)$ and $(d,d'x)$ reactions by a theoretical analysis using a quantum mechanical model, and to obtain a theoretical basis on the surface correction in the $(d,d'x)$ reaction.

\noindent
\textbf{Methods}:
The energy spectra and their radial distributions for the $(p,p'x)$ and $(d,d'x)$ reactions are calculated by the one-step semiclassical distorted wave model.

\noindent
\textbf{Results}:
The radial distribution of the energy spectra for the $(d,d'x)$ reaction is shifted toward the outer region of the nucleus compared to the $(p,p'x)$ reaction. 
Based on this finding, we consider a larger surface correction into a phenomenological model for the $(d,d'x)$ reaction than that for the $(p,p'x)$ reaction, and calculated values reproduce the experimental $(d,d'x)$ spectra well.

\noindent
\textbf{Conclusion}:
The peripherality of the $(d,d'x)$ reaction is more prominent than that of the $(p,p'x)$ reaction. 
The stronger surface correction thus should be introduced for the $(d,d'x)$ reaction than for the $(p,p'x)$ reaction.
\end{abstract}


\maketitle


\section{INTRODUCTION}\label{Sec_1}
Phenomenological models have played an important role in the nuclear data evaluation for nucleon-induced reactions~\cite{iwamoto2023, brown2018, plompen2020, koning2019, chadwick1999}.
For the calculation of pre-equilibrium processes, which are prominent at incident energies above about 10 MeV, the two-component exciton model~\cite{kalbach1986} has been widely used.
This phenomenological model has achieved great success in combination with global parameterization based on analyses of a large number of experimental data for $(N,N'x)$ reactions~\cite{kalbach1985, kalbach2000, koning2004}.
It has been argued in these analyses that the correction related to surface localization is important in reproducing the shape of experimental energy spectrum~\cite{kalbach1985, kalbach2000}.
Note that we will refer to the above correction as “surface correction" in this paper even though the correction has been referred to as “surface effects” in previous literature including Refs.~\cite{kalbach1985, kalbach2000, koning2004}.
This is because the “surface effects” is actually interpreted as a correction within the phenomenological model, as explained below. 

The surface correction is a simple means to effectively incorporate the peripherality of the $(N,N'x)$ reaction into the exciton model by restricting a hole degree of freedom with respect to the excitation energy.
The original exciton model~\cite{Griffin1966} assumes that the hole states could extend to infinite depth.
This model was later modified by considering a finite well depth in the density of states~\cite{betak1976}.
The modification restricts the region where the hole states are generated to a more realistic range.
In the surface correction, the finite well depth is made shallower only in the one-hole state, i.e., the state where the number of collision is small as discussed in Ref.~\cite{kalbach1985}.
This is an idea based on the following two matters in the exciton model that does not have a radial dependence: 
(i) reactions with fewer collisions would be more likely to occur around the nuclear surface,
(ii) nuclear potential is shallower around the nuclear surface than the interior.
It is worth mentioning that in Refs.~\cite{Avrigeanu1996, Avrigeanu1997}, it is indicated that the first $NN$ collision in $(N,N'x)$ reactions is localized on the nuclear surface using a semiclassical approach. Furthermore, the peripherality of the $(p,p'x)$ reaction is clarified by the theoretical analysis~\cite{watanabe1999} using the semiclassical distorted wave (SCDW) model, which has no free adjustable parameter~\cite{luo1991, kawai1992, ogata1999, weili1999, ogata2002}.

On another front, evaluation activities of deuteron nuclear data have gradually begun~\cite{iwamoto2023, nakayama2021, koning2019}, 
mainly motivated by the development of deuteron accelerator-based intensive neutron sources~\cite{Moeslang2006, ledoux2014}.
In the evaluation of deuteron nuclear data, phenomenological models~\cite{nakayama2022, kalbach2005} using state densities similar to those in the exciton model are often employed to calculate the components of direct inelastic scattering to continuum states.
These components correspond to inelastic scattering processes involving excitation of a nucleon particle-hole pair in the target nucleus.
The evaluation of the inelastic scattering is important since it affects the transport of deuterons in a material.
Therefore, as is the case with the exciton model, whether the surface correction should be taken into account in the state densities used in the above inelastic scattering model, and if so, to what extent, is an important factor in the deuteron nuclear data evaluation.
However, since experimental data for the $(d,d'x)$ reactions are very limited, there is concern that making this judgment based on a comparison with specific experimental data may give unphysical results under other conditions.
A theoretical basis for the surface correction of the $(d,d'x)$ reaction founded on a deep and comprehensive understanding of the peripherality of the reaction is required.

Under these circumstances, in Ref.~\cite{nakada2023}, the one-step SCDW model~\cite{luo1991} has been successfully applied to describe the inclusive $(d,d'x)$ reactions. This SCDW model has two noteworthy features. First, the model possesses the property of representing the inclusive cross section as an incoherent integral of contributions at individual collision points~\cite{cowley2000, watanabe1999}. This property allows us to analyze the peripherality of the inclusive reaction, i.e., to investigate where and to what extent the reaction occurs in the nucleus. The second is the adoption of local Fermi gas (LFG) to the single-particle states. Although LFG would be unrealistic for describing a specific nuclear state, it can reasonably describe the overall response of a nucleus, involving many initial and final states. Note that, in Ref~\cite{weili1999}, the SCDW model adopts the Wigner transform of one-body density matrices calculated with a single-particle state model for nuclei instead of LFG. On the other hand, we use LFG for reducing numerical tasks in this work.

The purpose of this study is to investigate the peripherality in the inclusive $(d,d'x)$ reaction and the cause of the difference from that in the inclusive $(p,p'x)$ reaction. 
The analyses are performed utilizing the above-mentioned property of the SCDW model.
Exploring the difference in the peripherality of the two reactions is of interest from the viewpoint not only of nuclear data evaluation but also of fundamental nuclear physics.
Furthermore, we apply the findings from the SCDW analysis to the implementation of the surface correction into the phenomenological deuteron inelastic scattering model.

The construction of this paper is as follows. 
In Sec.~\ref{Sec_2} we briefly describe the one-step SCDW model for the inclusive $(p,p'x)$ and $(d,d'x)$ reactions. 
We also explain the phenomenological inelastic scattering model. 
In Sec.~\ref{Sec_3} we compare the calculated energy spectrum of the inclusive $(p,p'x)$ and $(d,d'x)$ reactions with experimental data and clarify the difference of the peripherality between the two reactions. 
Improvements of the phenomenological model are also suggested.
Finally, a summary is given in Sec.~\ref{Sec_4}.

\section{Methods}\label{Sec_2}
\subsection{SCDW model}
\label{scdw_model}
We briefly describe the inclusive $(p,p'x)$ and $(d,d'x)$ reactions with the one-step SCDW model. The double differential cross section for the energy $E_{f}$ and the solid angle $\Omega_{f}$ of the emitted particle $c~(=~p~\mathrm{or}~d)$ is expressed with~\cite{nakada2023}
\begin{align}   
   \frac{\partial^2\sigma_{c}}{\partial E_{f}\partial\Omega_{f}}&=\left[\frac{A_{c}A}{A_{c}+A}\right]^2\frac{k_{f}}{k_{i}}\int d\bm{R}\nonumber\\
   &\times|\chi^{(-)}_{f}(\bm{R})|^{2}|\chi^{(+)}_{i}(\bm{R})|^{2}\left[\frac{\partial^2\sigma_{c}}{\partial E_{f}\partial\Omega_{f}}\right]_{\bm{R}}\rho(\bm{R}),
\label{eq_DDX}
\end{align}
 where $A_{c}$ and $A$ are the mass numbers of the particle $c$ and the target nucleus, respectively. $k_{i}~(k_{f})$ is the asymptotic momentum of the incident (emitted) particle, $\bm{R}$ is the coordinate of the collision point with respect to the center of the target nucleus. The distorted waves for $c$ in the initial and final states are denoted by $\chi_{i}$ and $\chi_{f}$, respectively. $\rho(\bm{R})$ is the nuclear density at $\bm{R}$. The averaged double differential cross section of the elementary process at $\bm{R}$ is given by
\begin{align}
  \left[\frac{\partial^2\sigma_{c}}{\partial E_{f}\partial\Omega_{f}}\right]_{\bm{R}}&=\frac{1}{(4\pi/3)k^3_{F}(\bm{R})}\left[\frac{A_{c}+1}{A_{c}}\right]^2\nonumber\\
  &\times\int_{k_{\alpha}\leq k_{F}(\bm{R})}d\bm{k}_{\alpha}\left(\frac{d\sigma_{cN}}{d\Omega}\right)_{\theta_{cN}(\bm{R}),E_{cN}(\bm{R})}\nonumber\\
  &\times\delta(E_{i}+\varepsilon_{\alpha}-E_{f}-\varepsilon_{\beta}),
\label{eq_DDX_ave}
\end{align}
where $\bm{k}_{\alpha}$ is the momentum of the nucleon of the target in the initial state. $k_{F}(\bm{R})$ represents the local Fermi momenta of nucleon. $d\sigma_{cN}/d\Omega$ is the free scattering cross section determined by the local scattering angle $\theta _{cN}(\bm{R})$ and the local scattering energy $E_{cN}(\bm{R})$ between $c$ and the nucleon of the target. The incident energy of $c$ is represented by $E_{i}$ and $\varepsilon_{\alpha}~(\varepsilon_{\beta})$ is the kinetic energy of the nucleon of the target nucleus in the initial (final) state. 
Equations \eqref{eq_DDX} and \eqref{eq_DDX_ave} are the same as Eqs.~(19) and (20) of Ref.~\cite{nakada2023}, respectively, and the derivation of the equations is discussed in detail in Ref.~\cite{nakada2023}.

From Eq.~\eqref{eq_DDX}, one can see that the inclusive cross section is described by an incoherent integral of $\bm{R}$. 
This description is made possible by the short-ranged property
of the kernel, Eq.~(2.8) of Ref.~\cite{luo1991}, which is realized when
a large number of s.p. states are involved; see Fig.~11 of
Ref.~\cite{luo1991}. In other words, the interference between the transition through different interacting points disappears in the situation. Thus, the projectile-nucleus cross section is given by integrating a product of the probability of the projectile reaching the point ${\bm R}$, that of the ejectile being emitted from ${\bm R}$, and the averaged cross section of the elementary process at ${\bm R}$. For more details, readers are referred to Ref.~\cite{luo1991}.

By applying the local Fermi gas model (LFG) to the initial and final nuclear single-particle states, $k_F(\bm{R})$ is given by $k_F(\bm{R})=[3\pi^2\rho(\bm{R})/2]^{1/3}$~\cite{luo1991}. Therefore, LFG can takes into account the spread of nucleon density on the nuclear surface. To clearly show the contribution of the nuclear surface to the inclusive cross section, we also consider the Fermi gas model (FG), which gives a uniform Fermi momentum distribution. When FG is used instead of LFG, $k_{F}(\bm{R})$ becomes step functions with respect to $R$ in Eq.~\eqref{eq_DDX_ave}. On the other hand, the Fermi momentum in LFG decreases smoothly as $R$ increases, as shown later in Fig.~\ref{fig:kF}. 

The energy spectra of the $(p,p'x)$ and $(d,d'x)$ reactions are obtained by integrating Eq.~\eqref{eq_DDX} over $\Omega_f$:
\begin{align}   
   \frac{d\sigma_c}{dE_f}=\int d\Omega_f \frac{\partial^2\sigma_{c}}{\partial E_{f}\partial\Omega_{f}}.
\label{eq_DX}
\end{align}

We use the proton- and deuteron-nucleus global optical potential by Koning-Delaroche~\cite{koning2003} and An-Cai~\cite{An2006} for describing projectile-nucleus scattering, respectively. 
The nuclear density $\rho(R)$ is assumed to be the Woods-Saxon form, where the radial parameter is defined as $R_{\rho} = r_{\rho}A^{1/3}$, with $r_{\rho}$ = 1.15 fm, 
and the diffuseness parameter is set to $a_\rho = 0.5$ fm as in Ref.~\cite{nakada2023}. 
For the differential cross sections of $d$-$N$ scattering used in the $(d,d'x)$ calculation, we utilize the numerical table from Ref.~\cite{chazono2022a}, which was made by fitting experimental data of $p$-$d$ elastic scattering. 
In the calculations of the $(p,p'x)$ process, the free $p$-$N$ scattering cross sections are calculated by using the nucleon-nucleon $t$ matrix provided by Franey and Love~\cite{Love1981,Franey1985}.

\subsection{Phenomenological model by Kalbach}
\label{Kalbach_model}
As a phenomenological model for continuum deuteron inelastic scattering, we adopt the model proposed by Kalbach~\cite{kalbach1977}.
Although this model was originally developed for $\alpha$-particle-induced reactions, the model itself is considered to be applicable to deuteron-induced ones~\cite{nakayama2022}.
Note that what is calculated in this model is the direct process component, which is not included in the exciton model.

According to Ref.~\cite{kalbach1977}, the energy spectrum of the direct deuteron inelastic scattering with a nucleon particle-hole pair creation is calculated by:
\begin{equation}
\begin{split}
		\frac{d\sigma_d}{dE_f} = C\frac{\sigma_{\mathrm{rea}}(E_i)}{E^3_i}(2s+1) E_f 
           &\sigma_{\mathrm{rea}}(E_f) \frac{D_{F}(U)}{A^2}
        \\ &\times 0.12\ \left(\frac{\mathrm{MeV}^{2}}{\mathrm{mb}}\right),
\label{eq_kalbach_dx}
\end{split}
\end{equation}
where $C$ is a normalization constant determined from fitting experimental energy spectrum, $\sigma_{\mathrm{rea}}(E)$ is the deuteron total reaction cross section at scattering energy $E$, and $E_i$ and $E_f$ are the incident and emitted deuteron energies, respectively.
According to Ref.~\cite{kalbach1977}, the overall normalization constant $C$ can vary in the range of a factor of 3 depending on the target.
As shown later in Sec.~\ref{app_to_phenom}, in this study, $C$ is adjusted so that the peak of the spectrum matches available experimental data.
$\sigma_{\mathrm{rea}}(E)$ is obtained from the optical model calculation with the global potential by An-Cai~\cite{An2006}.
$s$ is the spin of the emitted deuteron.
$A$ denotes the mass number of the target, and $U$ represents the effective excitation energy of residual nucleus considering the paring effect~\cite{Kalbach1995}.
The final state density $D_{F}$ is expressed as follows:
\begin{equation}
		D_{F}(U) = (g_{n}^{2}U+g_{p}^{2}U)f(V,U),
\label{eq_state_density}
\end{equation}
where $g_n=N/(13~\mathrm{MeV})$, $g_p=Z/(13~\mathrm{MeV})$, and $N$ and $Z$ are the numbers of neutrons and protons in the target.
The setting of values for $g_n$ and $g_p$ follows the original paper by Kalbach~\cite{kalbach1977}.
The squaring of $g_n$ and $g_p$ corresponds to considering the $1p$-$1h$ states of nucleons.
The finite well function $f(V,U)$ with the well depth $V$ is defined for the $1p$-$1h$ states as follows~\cite{betak1976}:
\begin{equation}
		f(V,U) = 1-\left[\frac{U-V}{U}\right] \Theta(U-V),
\label{eq_finite_depth}
\end{equation}
where $\Theta$ is the unit step function. 
Note that the finite well function is not taken into account in the original Kalbach model ~\cite{kalbach1977} and the first application of the model to the deuteron-induced reaction~\cite{nakayama2022}.
The introduction of the finite well function enables us to consider the surface correction as shown below.

As mentioned in Sec.~\ref{Sec_1}, a simple method to include the peripherality of the nuclear reaction in the exciton model was proposed in Ref.~\cite{kalbach1985}, and has been successfully applied to the analysis of $(N,N'x)$ reactions~\cite{kalbach2000, koning2004}.
In the method, the peripherality is effectively taken into account by fixing the small value of $V$ for the $h$=1 state only.
In this work we apply this method to Eq.~(\ref{eq_kalbach_dx}).
In other words, we set $V$ in Eq.~(\ref{eq_finite_depth}) smaller than the typical value of 38 MeV associated to the Fermi energy.
By introducing $F(U,V), D_F(U)$ is replaced by smaller density of states $D_F(V)$ when $U > V$.
Therefore, giving a smaller V corresponds to making the reaction more likely occur in the peripheral region of the nucleus, based on the fact that the nuclear potential is shallower around the nuclear surface than in the interior.
The details of fixing the value of $V$ are described later in Sec.~\ref{app_to_phenom}.

Following the idea in Kalbach's original paper~\cite{kalbach1977}, the values calculated with the inelastic scattering model are treated as the direct component in this study. 
Therefore, we do not consider in the model the reduction of the composite nucleus production cross section caused by other direct processes, such as the breakup reactions.
The use of the total reaction cross section obtained from the optical model as $\sigma_{\mathrm{rea}}$ in Eq.~(\ref{eq_kalbach_dx}) reflects the above idea.
The interference with other direct processes is not considered.

\section{RESULTS AND DISCUSSION}\label{Sec_3}
\subsection{Influence of nuclear surface on energy spectrum}
\label{comp_LFG_FG}

\begin{figure}[h]
    \centering
    \includegraphics[width=0.90\hsize]{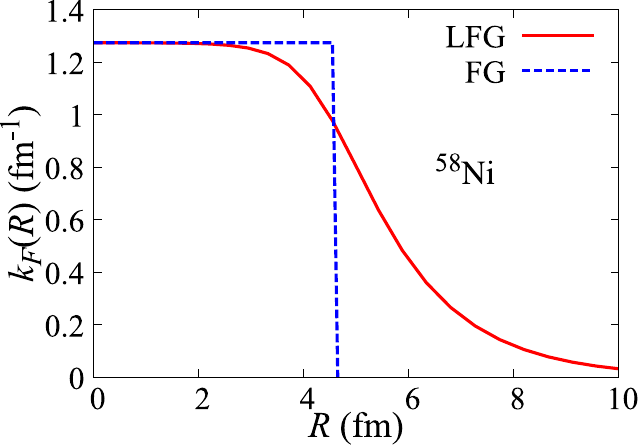}
    \caption{The $R$-dependence of the Fermi momenta of $^{58}\mathrm{Ni}$. The solid (dashed) line represents the Fermi momentum with LFG (FG).}
\label{fig:kF}
\end{figure}

We compare the energy spectra of inclusive $(p,p'x)$ and $(d,d'x)$ reactions calculated with SCDW using LFG and FG, and experimental data. 
Figure~\ref{fig:kF} shows the radial distributions of the Fermi momenta of $^{58}\mathrm{Ni}$ calculated with LFG (solid line) and FG (dashed line).

In FG, no reaction is allowed in $R\gtrsim4.4$ fm because $k_F(R)$ is zero in the region. 
On the other hand, in LFG, reactions are allowed beyond the boundary because LFG incorporates the diffuseness around the nuclear surface. 
Therefore, the difference between the SCDW calculation with LFG and that with FG can be tied to the influence of considering the nuclear surface.
\begin{figure*}
    \centering
    \includegraphics[width=0.95\hsize]{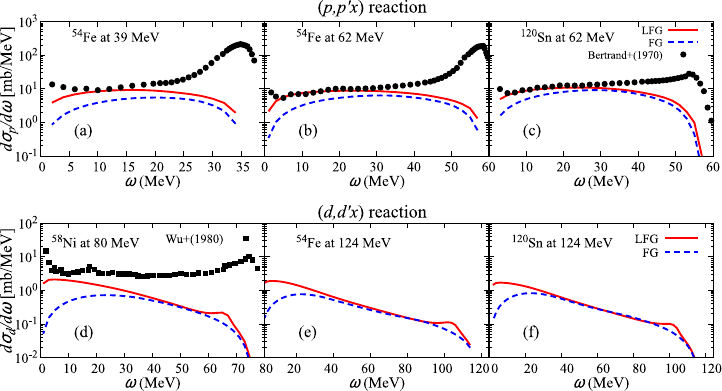}
    \caption{Comparison of the calculated energy spectra and the experimental data~\cite{bertrand1973} of the $(p,p'x)$ reaction on $^{54}\mathrm{Fe}$ at (a) 39 MeV and (b) 62 MeV and (c) $^{120}\mathrm{Sn}$ at 62 MeV. (d) The energy spectrum and the experimental data~\cite{wu1979} of the $(d,d'x)$ reaction on $^{58}\mathrm{Ni}$ at 80 MeV.  (e) and (f) are same as (b) and (c) but for the $(d,d'x)$ reaction at 124 MeV. The horizontal axis represents the energy transfer $\omega \equiv E_i-E_f$. The solid and dashed lines represent the calculated results with LFG and FG, respectively.}
\label{fig:DX}
\end{figure*}
Figure~\ref{fig:DX} shows the energy spectra of the $(p,p'x)$ and $(d,d'x)$ reactions for several incident energies and targets. The horizontal axis is the energy transfer $\omega \equiv E_i-E_f$. The solid (dashed) lines are calculated results with LFG (FG).
The experimental data are taken from Refs.~\cite{bertrand1973,wu1979}. 
In order to make the velocity of projectile same, the incident energies per nucleon in Fig.~\ref{fig:DX}(e) and (f) are the same as those in (b) and (c), respectively. Note that these results are calculated with SCDW which has no free adjustable parameter.

For the $(p,p'x)$ reactions, the energy spectra calculated with SCDW using LFG reproduces the experimental data well except for the region with large energy transfer $\omega$ as shown in Fig~\ref{fig:DX}(a), (b), and (c).
For the $(d,d'x)$ reactions, the energy spectra calculated with LFG reasonably reproduce the experimental data in the region with $\omega\lesssim15$ MeV, as shown in Fig.~\ref{fig:DX}(d). 
In both reactions, the calculated energy spectrum undershoot the experimental data in the region with low emission energies.
This is because the contributions of particle emission from the multi-step direct process and the compound process are dominant in that region~\cite{cowley2000,nakayama2022}.
For this reason, in what follows, we will discuss the energy spectra in the region with $5\leq\omega\leq15$ MeV, where the one-step process is dominant and the elastic scattering events are not included.

By comparing the energy spectra calculated with LFG and FG, we can see that consideration of the nuclear surface is necessary to reproduce experimental data in both the $(p,p'x)$ and $(d,d'x)$ reactions. 
The importance of considering the nuclear surface does not depend so much on the incident energy and target nucleus in both reactions.
We can also find that the influence of the nuclear surface gets larger as $\omega$ decreases. This result implies that the peripherality of the reactions becomes stronger as $\omega$ decreases. 
Moreover, the $\omega$ dependence is more prominent in the $(d,d'x)$ reactions than the $(p,p'x)$ reactions.

\subsection{Difference in peripherality of $(p,p'x)$ and $(d,d'x)$ reactions}
\label{surface_effect}
To analyze the difference in the influence of nuclear surface between the $(p,p'x)$ and $(d,d'x)$ reactions in more detail, we discuss the radial distribution of the energy spectra.

The radial distribution of the energy spectra is defined by
\begin{align}
  f_c(\omega,R)&\equiv\left[\frac{A_{c}A}{A_{c}+A}\right]^2\frac{k_{f}}{k_{i}}\int d\Omega_f R^2\int d\Omega\nonumber\\
   &\times|\chi^{(-)}_{f}(\bm{R})|^{2}|\chi^{(+)}_{i}(\bm{R})|^{2}\left[\frac{\partial^2\sigma_{c}}{\partial E_{f}\partial\Omega_{f}}\right]_{\bm{R}}\rho(\bm{R}),
\label{eq_DX_R}
\end{align}
where $\Omega$ is the solid angle of $\bm{R}$. 
Note that $f_c(\omega,R)$ satisfies $d\sigma_c/dE_f=\int f_c(\omega,R)dR$.

\begin{figure}[h]
    \centering
    \includegraphics[width=0.95\hsize]{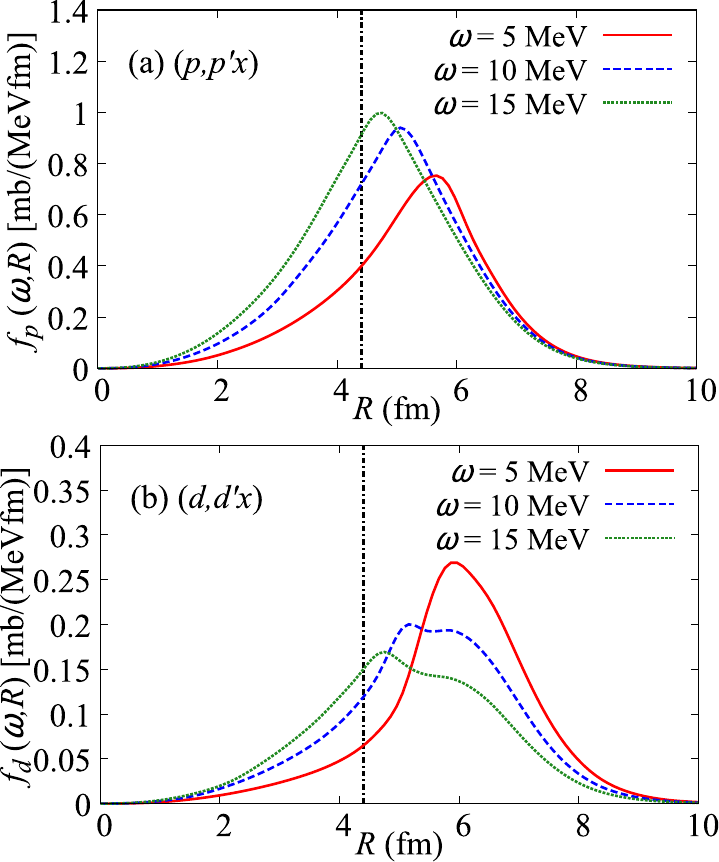}
    \caption{$f_c(\omega,R)$ on $^{54}\mathrm{Fe}$ at 62 MeV per nucleon as a function of $R$ in the (a) $(p,p'x)$ and (b) $(d,d'x)$ reactions. 
    The solid, dashed, and dotted lines show $f_c(\omega,R)$ with $\omega=$ 5, 10, and 15 MeV, respectively. 
    The vertical dash-dotted lines represent $R=$ 4.4 fm, where the Fermi momentum with FG of $^{54}\mathrm{Fe}$ becomes zero.}
\label{fig:DX_R}
\end{figure}

Figure~\ref{fig:DX_R} shows the $f_c(\omega,R)$ in the calculation with LFG for (a) the $^{54}\mathrm{Fe}(p,p'x)$ reaction at 62 MeV and (b) the $^{54}\mathrm{Fe}(d,d'x)$ reaction at 124 MeV (62 MeV per nucleon). 
The solid, dashed, and dotted lines represent $f_c(\omega,R)$ with $\omega=5,~10,~\mathrm{and}~15$ MeV, respectively. 
The vertical dash-dotted lines indicate $R\simeq4.4$ fm, where the Fermi momentum of $^{54}\mathrm{Fe}$ is zero in FG.

As shown in Fig.~\ref{fig:DX_R}, for a given $\omega$, the peak position is almost the same for the $(p,p'x)$ and $(d,d'x)$ reactions and the peak positions shift outwards as $\omega$ decreases. In contrast, the $\omega$ dependence of the peak heights of the two reactions is opposite. The peak height of the $(p,p'x)$ reaction becomes smaller as $\omega$ decreases, while that of the $(d,d'x)$ reaction becomes larger as $\omega$ decreases.

\begin{figure}[h]
    \centering
    \includegraphics[width=0.95\hsize]{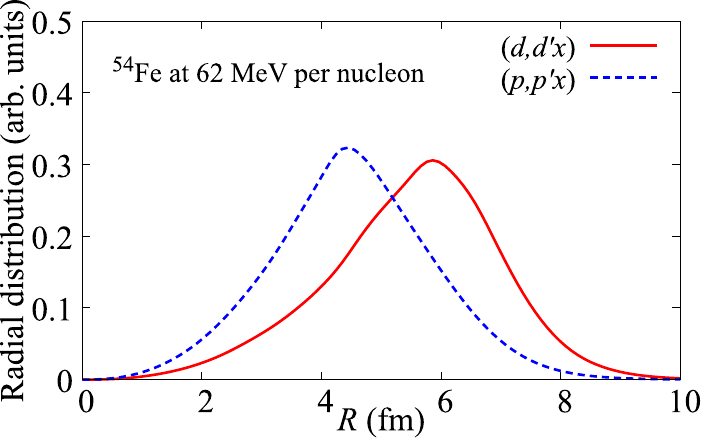}
    \caption{The radial distributions of the integration of $f_c(\omega, R)$ over $\omega$ in the range of 5 to 15 MeV. The solid and dashed lines show the results for $(d,d'x)$ and $(p,p'x)$ reactions, respectively. 
    Each distribution is normalized so that the integral value over $R$ is unity.}
\label{fig:DX_RTotal}
\end{figure}

To clearly show how this significant difference affects the peripherality of these reactions, we present in Fig.~\ref{fig:DX_RTotal} the integrated values of $f_c(\omega,R)$ in the range of 5 $\le$ $\omega$ $\le$ 15 MeV. 
The solid and dashed lines represent the calculated values of the $(d,d'x)$ and $(p,p'x)$ reactions, respectively. 
Each of them is normalized so that the integrated value over $R$ is unity. 
It is seen that the radial peak position is about 4.3 fm for the $(p,p'x)$ reaction.
On the other hand, for the $(d,d'x)$ reaction, the peak position is about 6.0 fm and a large proportion of the reaction occurs in the outer region of the nucleus. This result clearly shows that the $(d,d'x)$ reaction has a stronger peripherality than the $(p,p'x)$ reaction.

\begin{figure}[h]
    \centering
    \includegraphics[width=0.95\hsize]{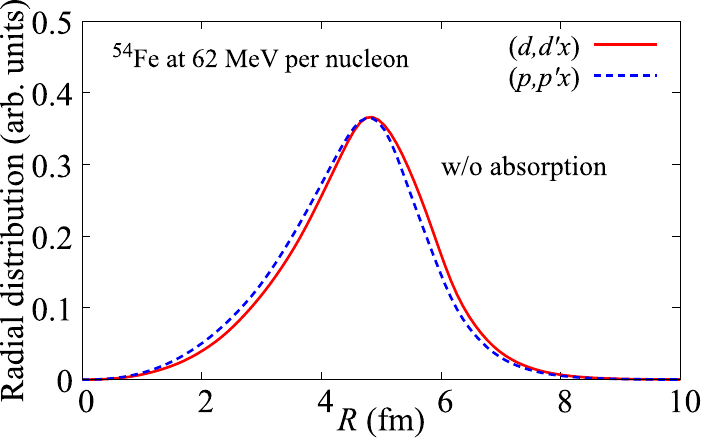}
    \caption{Same as Fig.~\ref{fig:DX_RTotal} but without the absorption by $^{54}\mathrm{Fe}$.}
\label{fig:DX_RTotal_ImV0}
\end{figure}

Next, we discuss the cause of the difference in the peripherality of the two reactions. 
Figure~\ref{fig:DX_RTotal_ImV0} is the same as Fig.~\ref{fig:DX_RTotal} but for the calculations ignoring the absorption of the protons and deuterons by the nucleus; the imaginary part of the optical potential is set to zero in the calculations.
When we ignore the absorption, the radial distributions of the two reactions are almost identical.
Therefore, we can see that the stronger peripherality of the $(d,d'x)$ reaction than that of the $(p,p'x)$ reaction is attributed to the strong absorption of deuteron by the nucleus. 
We can also confirm that in Fig~\ref{fig:DX_R}, the strong absorption of the deuteron suppresses the reaction in the inner region of the nucleus in the $(d,d'x)$ than in the $(p,p'x)$ reaction.

\subsection{Application to phenomenological model}
\label{app_to_phenom}
\begin{figure}[h]
    \centering
    \includegraphics[width=1.0\hsize]{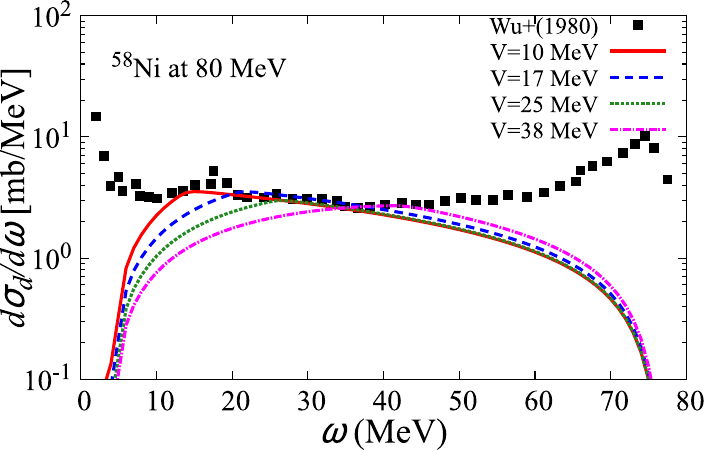}
    \caption{Comparison of the experimental and calculated energy spectrum of the inclusive $(d,d'x)$ reaction on $^{58}$Ni at 80 MeV.
    The solid line denotes the calculation using a well depth of 10 MeV optimized to fit the experimental spectrum.
    The dashed and dotted lines are the results with a well depth of 17 and 25 MeV obtained from the parameterization for proton-induced reactions in Refs.~\cite{koning2004,kalbach2000}, respectively.
    The dash-dotted line is the calculation with a typical well depth of 38 MeV, and this means the surface correction are not introduced.
    The horizontal axis represents the energy transfer $\omega$.
    The experimental data are taken from Ref.~\cite{wu1979}.
    }
    \label{fig:phenom}
\end{figure}

Next, we apply the findings obtained from the previous sections to the implementation of the surface correction into the phenomenological model.
Figure~\ref{fig:phenom} shows the comparison of the experimental data and the energy spectra calculated with the Kalbach model presented in Sec.~\ref{Kalbach_model}. 
The horizontal axis represents the energy transfer $\omega$.
The experimental data are obtained from Ref.~\cite{wu1979}.

Four lines are shown in the figure.
The solid line is the result of the calculation in which the well depth $V$ is optimized to reproduce the experimental spectrum.
The optimised value of $V$ is 10 MeV.
The normalization factor $C$ is adjusted so that the peak of the spectrum matches the experimental data. 
This adjustment holds true also for the three calculations below.
The dashed line shows the result of using 17 MeV as the value of $V$.
In Ref.~\cite{kalbach2000}, a value of 17 MeV was proposed as the optimum value of $V$ for the $(p,p'x)$ reaction, independent of the incident energy and target nucleus.
A local optimum value of $V=12(+3/-4)$ MeV has also been derived in Ref.~\cite{kalbach2000}.
On the other hand, it is also discussed in Ref.~\cite{kalbach2000} that the value of $V$ can vary depending on the background of the experiments being compared.
We thus adopted the global optimum value of $V=17$ MeV, obtained from the analysis of several tens of experiments, as one of the targets for comparison.
The dotted line is the result with the parameterization optimized for proton-induced reactions derived in Ref.~\cite{koning2004}.
In Ref.~\cite{koning2004}, the analyses using the exciton model were performed for $(N,N'x)$ reactions up to 200 MeV for various targets, and the different parameterizations for neutron- and proton-induced reactions were given based on the results.
The value of $V$ in the case of proton-induced reactions is given as follows:
\begin{equation}
		V = 22 + 16 \frac{E^4}{E^4 + (450/A^{1/3})^4}\:\mathrm{MeV},
\label{eq_v_proton}
\end{equation}
where $E$ and $A$ are the incident energy and the target mass, respectively.
In the case shown in Fig.~\ref{fig:phenom}, Eq.~(\ref{eq_v_proton}) gives about 25 MeV as the value of $V$.
On the other hand, 
Finally, the dash-dotted line is the calculation with a typical well depth of 38 MeV.
Using this value of $V$ corresponds to not introducing the surface correction.

Note that in Ref.~\cite{nakayama2022} we have performed an analysis of the double differential cross sections for the same target and incident energy at 20\textdegree.
At such a forward angle, the components corresponding to giant resonance states were dominant in the region of $10\lesssim\omega\lesssim20$ MeV.
On the other hand, the use of angle-integrated DX as in Fig.~\ref{fig:phenom} makes it easier to discuss the direct inelastic scattering component calculated with the Kalbach model, since the giant resonance components make up a smaller fraction of the total in the above $\omega$ region.

As shown in the figure, the calculation not introducing the surface correction does not reproduce the shape of the experimental spectrum. 
Note that the components above 50 MeV are mainly due to the pre-equilibrium process and the compound nucleus process.
The large component below a few MeV is attributed to inelastic scattering to low-lying discrete levels and elastic scattering.
These components are not considered in the Kalbach model and are thus outside the scope of the present discussion.

In terms of the calculations incorporating the surface correction, the use of the optimized values for proton-induced reactions improves the agreement with experimental data, but underestimation in the high-energy region is still seen.
On the other hand, when the value of $V$ is more decreased, the calculation results reproduce the experimental energy spectrum better over almost the entire energy range to be considered in the Kalbach model.
As discussed in Sec.~\ref{Kalbach_model}, the smaller value of $V$ makes the reaction more likely occur in the peripheral region of the nucleus. Therefore, this result is consistent with the results of the present SCDW analysis.
In other words, parameter optimization of the phenomenological Kalbach model has been justified also by theoretical analysis with the SCDW model.

In Ref.~\cite{kalbach1985}, it is argued that the magnitude of the surface correction can also vary depending on what excitation energy dependence is assumed for the single-particle level density.
Meanwhile, an equispacing model that does not assume an excitation energy dependence of the single-particle level density is adopted in Refs.~\cite{koning2004, kalbach2000}, and the present study also uses the same equispacing model as presented in Eq.~(\ref{eq_state_density}).
It is worth mentioning that these comparisons thus are valid as a discussion on the surface correction.

\section{SUMMARY}\label{Sec_4}
We have clarified the difference in the peripherality of the inclusive $(p,p'x)$ and $(d,d'x)$ reactions and the cause of the difference. 
The energy spectra calculated by SCDW with LFG and FG were compared to the experimental data of the $(p,p'x)$ and $(d,d'x)$ reactions on several combinations of targets and incident energies. 
By comparing the energy spectra with LFG and FG, we have found that the consideration of the nuclear surface is necessary to reproduce the experimental data for both $(p,p'x)$ and $(d,d'x)$ reactions. 

We have investigated the difference in the peripherality of the $(p,p'x)$ and $(d,d'x)$ reactions by comparing the radial distributions of the energy spectra. 
The radial peak positions for each energy transfer $\omega$ shifts to the outer region of the nucleus as $\omega$ decrease and the tendencies were almost the same for the two reactions.
However, the opposite trend was observed between the two reactions in terms of the peak height.
As a result, it has been found that the peripherality get stronger for the $(d,d'x)$ reaction than the $(p,p'x)$ reaction.
Moreover, we compared the radial distributions ignoring the absorption by the nucleus.
The results show almost identical radial distributions and it has been found that the cause of the stronger peripherality of the $(d,d'x)$ reaction is the stronger absorption of deuteron than proton.

We have applied the above findings on the peripherality of $(d,d'x)$ reaction to the improvement of the phenomenological model by Kalbach to calculate continuum deuteron inelastic scattering.
For the finite well depth $V$, which is the adjustable parameter associated with the surface correction, the optimized value for the proton-induced reactions was insufficient to reproduce the experimental data of the $(d,d'x)$ reaction.
When the value of $V$ was set much smaller, as suggested by the analysis with SCDW model, the calculated values reproduced the experimental data well over a wide emission energy range.
This result has indicated that parameter optimization of the phenomenological reaction model has been justified by the SCDW model.

As one of the future works, it will be necessary to extend the present one-step SCDW model to describe the multi-step processes.
With this extension, the SCDW model could guide the improvement of phenomenological models not only with respect to surface corrections, which are prominent in the small $\omega$ region, but also with respect to wider energy spectra and angular distributions of outgoing deuterons.
On another front, it is of interest to perform a similar analysis for other particles. 
Ref.~\cite{watanabe1999} was the first example to analyze the peripherality of the $(p,p'x)$ reaction in detail using the SCDW model. In this study, we presented that a similar analysis is also valid for reactions induced by deuteron. Future similar analyses for various composite particles will improve our understanding of the peripherality of nuclear reactions. 
$\alpha$-particle is a good candidate since it is well known to undergo strong absorption and is of large importance in the application fields.
\begin{acknowledgments}
The authors thank Y. Chazono for providing us with the table of the $d$-$N$ scattering cross section, and thank S. Kawase for fruitful discussions. This work has been supported in part by Grants-in-Aid of the Japan Society for the Promotion of Science (Grants No. JP20K14475, No. JP21H00125, and No. JP21H04975). The computation was carried out with the computer facilities at the Research Center for Nuclear Physics, Osaka University.
\end{acknowledgments}



\bibliographystyle{apsrev4-2}
\bibliography{prc2}

\end{document}